\documentclass[aps,prb,twocolumn,superscriptaddress,showpacs]{revtex4}

\usepackage{graphicx}
\usepackage{dcolumn}
\usepackage{bm}

\begin{document}

\title{Thermal conductivity of pure and Zn-doped LiCu$_2$O$_2$ single crystals}

\author{X. G. Liu}
\affiliation{Hefei National Laboratory for Physical Sciences at
Microscale, University of Science and Technology of China, Hefei,
Anhui 230026, People's Republic of China}

\author{X. M. Wang}
\affiliation{Hefei National Laboratory for Physical Sciences at
Microscale, University of Science and Technology of China, Hefei,
Anhui 230026, People's Republic of China}

\author{W. P. Ke}
\affiliation{Hefei National Laboratory for Physical Sciences at
Microscale, University of Science and Technology of China, Hefei,
Anhui 230026, People's Republic of China}

\author{W. Tao}
\affiliation{Hefei National Laboratory for Physical Sciences at
Microscale, University of Science and Technology of China, Hefei,
Anhui 230026, People's Republic of China}

\author{X. Zhao}
\affiliation{School of Physical Sciences, University of Science
and Technology of China, Hefei, Anhui 230026, People's Republic of
China}

\author{X. F. Sun}
\email{xfsun@ustc.edu.cn}

\affiliation{Hefei National Laboratory for Physical Sciences at
Microscale, University of Science and Technology of China, Hefei,
Anhui 230026, People's Republic of China}

\date{\today}

\begin{abstract}

We report a study of the low-temperature thermal conductivity
($\kappa$) of pure and Zn-doped LiCu$_2$O$_2$ single crystals. The
$\kappa(T)$ of pure LiCu$_2$O$_2$ single crystal shows a
double-peak behavior, with two peaks locating at 48 K and 14 K,
respectively. The different dependences of the peaks on the Zn
concentration indicate that the high-$T$ peak is likely due to the
phonon transport while the low-$T$ one is attributed to the magnon
transport in the spin spiral ordering state. In addition, the
magnetic field can gradually suppress the low-$T$ peak but does
not affect the high-$T$ one; this further confirms that the
low-$T$ peak is originated from the magnon heat transport.

\end{abstract}


\maketitle

\section{Introduction}

LiCu$_2$O$_2$ is the first example of Cu-based multiferroic
material and is particularly attractive because of its
one-dimensional spin structure.\cite{Park, Masuda, Seki, Huang,
Rusydi, Capogna, Moskvin, Mihaly, Huvonen, Hsu1, Hsu2} It promises
a new routine to find multiferroicity in some low-dimensional
quantum magnets that exhibit the magnetic frustration
effect.\cite{Naito, Lawes, Kagawa} It is known that LiCu$_2$O$_2$
crystallizes in an orthorhombic unit cell with space group $Pnma$
and lattice parameters $a$ = 5.734(4) {\AA}, $b$ = 2.856(2) {\AA}
and $c$ = 12.415(6) {\AA} at room temperature.\cite{Park} There
are an equal number of Cu$^+$ and Cu$^{2+}$ ions in distinctly
nonequivalent crystallographic positions, only the latter of which
carry spin $S$ = 1/2. The Cu$^{2+}$ ions are sitting on the center
of edge-sharing CuO$_4$ plaquettes and form edge-shared chains
running along the $b$ axis with the Cu-O-Cu bond angle of
94$^\circ$.\cite{Park} The competition of the nearest-neighboring
ferromagnetic (FM) interaction and the next-nearest-neighboring
antiferromagnetic (AF) interaction of Cu$^{2+}$ spins leads to
magnetic frustration and a spiral (helicoidal) magnetic order
below $\sim$ 24 K.\cite{Park} More exactly, two AF transitions
were found at $T_{N1}$ = 24.6 K and $T_{N2}$ = 23.2 K with a
sinusoidal spin order at $T_{N2} < T < T_{N1}$ and an
incommensurate cycloidal spin order at $T < T_{N2}$.\cite{Seki,
Huang, Rusydi, Capogna, Moskvin} A spontaneous polarization along
the $c$ axis emerges at the second phase transition\cite{Park} and
was discussed to be originated from the inverse DM interaction of
neighboring spins or the nonrelativistic exchange.\cite{Seki,
Moskvin} Furthermore, the magnetic field applied along the $b$
axis leads to the Cu$^{2+}$ spin spiral plane flipping from the
$bc$ to the $ab$ plane and consequently results in the flip of the
polarization from the $c$ to the $a$ direction.\cite{Park}

The low-temperature thermal conductivity is an effective probe for
studying the transport properties of phonons and magnetic
excitations, which might be of particularly interesting for the
low-dimensional quantum magnets.\cite{Brenig1, Hess1,
Sologubenko1} It is well known that the elementary excitations in
magnetic long-range ordered states, magnons, can contribute to the
heat transport properties by acting as either heat carriers or
phonon scatterers.\cite{Berman} In low-dimensional spin systems,
even without exhibiting long-range ordering, the magnetic
excitations can effectively transport heat because of the strong
quantum fluctuations.\cite{Brenig1} For example, the recent
experiments revealed extremely large magnetic heat conductivity in
one-dimensional spin 1/2 systems, such as SrCuO$_2$,
Sr$_2$CuO$_3$, CaCu$_2$O$_3$ and
(Sr,Ca,La)$_{14}$Cu$_{24}$O$_{41}$.\cite{Sologubenko2, Kawamata,
Hess2, Hess3, Hlubek} These results were theoretically well
understood as the ballistic transport of spinons or magnons. On
the other hand, the thermal conductivity can effectively detect
the transitions of magnetic structure like spin flop,
reorientation or polarization.\cite{Spin_flop} In particular, the
low-$T$ thermal conductivity was found to display drastic changes
across these kinds of transition in some other families of
multiferroic materials, for example, HoMnO$_3$ and
GdFeO$_3$.\cite{Wang, Zhao}

Although the low dimensionality of the magnetic structure
LiCu$_2$O$_2$ has already known for long time, the heat transport
properties have not been investigated for this material. In this
work, we study the temperature and magnetic-field dependences of
thermal conductivity ($\kappa$) of LiCu$_2$O$_2$ single crystals
for probing the nature of magnetic excitations and the coupling
between spin and lattice. It is found that the $\kappa(T)$ data
show two peaks at 48 K and 14 K, which are above and below the
long-range magnetic transition temperatures, respectively. The Zn
substitution for Cu is found to be able to effectively suppress
the two peaks but show quite different doping dependences. The
magnetic field applied in the $ab$ plane only suppresses the
low-$T$ peak. These results manifest that the low-$T$ peak is
likely due to the magnons contribution to the heat transport
acting as heat carriers, while the high-$T$ peak is phonon peak.

\section{Experiments}

High-quality single crystals of LiCu$_{2-x}$Zn$_x$O$_2$ with the
nominal compositions $x$ = 0, 0.02, 0.04, 0.10 and 0.20 are grown
by a self-flux method. Correspondingly, the actual Zn contents of
these crystals are $x$ = 0, 0.013, 0.027, 0.064 and 0.177,
measured by the inductively-coupled plasma atomic-emission
spectroscopy (ICP-AES) (with 10\% uncertainty of measurements).
The as-grown LiCu$_2$O$_2$ single crystals have plate-like shape
with typical size as large as 10 $\times$ 8 $\times$ 0.5 mm$^3$.
Upon doping Zn, the sizes of crystals decrease gradually to about
5 $\times$ 3 $\times$ 0.5 mm$^3$ for $x$ = 0.177. The quality of
the crystals, judged from the x-ray diffraction results, does not
decay significantly. The largest surfaces of single crystals are
confirmed to be the $ab$ planes by the x-ray diffraction and Laue
back reflection.

The magnetic susceptibility ($\chi$) is measured using a
superconducting quantum interference device magnetometer (SQUID,
Quantum Design). The specific heat is measured by the relaxation
method in a physical property measurement system (PPMS, Quantum
Design) from 2 to 300 K. The in-plane thermal conductivity is
measured in PPMS with ``one heater, two thermometers"
configuration or in a $^4$He cryostat using a Chromel-Constantan
thermocouple.\cite{Wang, Zhao, Sun_DTN}

\section{Results and Discussion}

\begin{figure}
\includegraphics[clip,width=8.5cm]{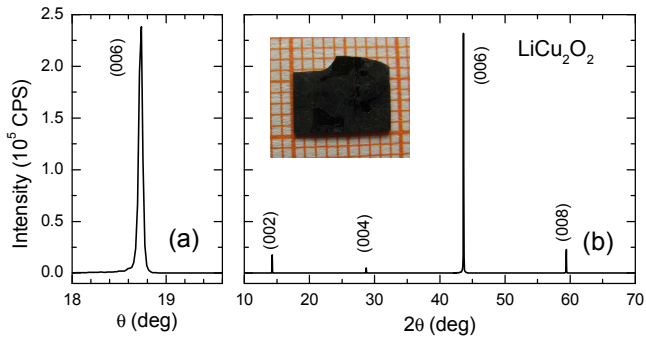}
\caption{(Color online) (a) X-ray rocking curve of the (006)
reflection of a LiCu$_2$O$_2$ single crystal. The FWHM of the peak
is only 0.05$^\circ$. (b) X-ray $(00l)$ diffraction pattern of the
same crystal. Inset: The photograph of this crystal.}
\end{figure}

LiCu$_2$O$_2$ single crystals have plate-like shape and shinning
surfaces. Using x-ray diffraction, it is found that the large
surface of the crystals are the $ab$ plane, so it is easy to get
the $(00l)$ diffraction pattern. Figure 1 shows the x-ray
diffraction results of a representative LiCu$_2$O$_2$ single
crystal, including the $(00l)$ diffraction pattern and the rocking
curve of (006) peak. Figure 1(b) does not show any sign of other
phases, indicating the high purity of this sample. The full width
at half maximum (FWHM) of the (006) reflection is as small as
0.05$^\circ$, indicating the perfect crystallinity of this sample.
X-ray Laue back reflection is also used to determine the
crystallographic axes. A fine twin structure of the $ab$ plane is
found, which is in agreement with the previous observation through
a polarized optical microscope.\cite{Park} The origin is that the
$a$-axis lattice constant is nearly the twice of the $b$-axis
length. Because of the existence of twin structure, it is
impossible to measure the in-plane anisotropy of the transport
properties of LiCu$_2$O$_2$ single crystals, although all the
samples are cut along the $a$ (or $b$) axis.

\begin{figure}
\includegraphics[clip,width=7.5cm]{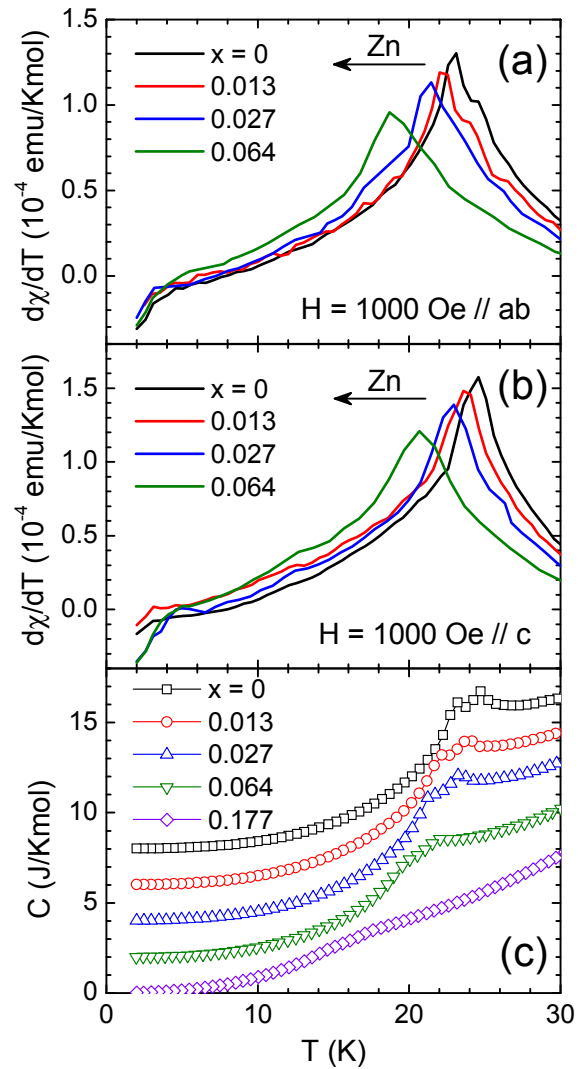}
\caption{(Color online) (a,b) Temperature dependences of
$d\chi/dT$ for LiCu$_{2-x}$Zn$_x$O$_2$ single crystals along the
$ab$ plane and the $c$ axis measured with 1000 Oe magnetic field.
(c) Low-temperature specific heat of LiCu$_{2-x}$Zn$_x$O$_2$
single crystals. For clarity, the data are shifted upward by 2
J/Kmol one by one.}
\end{figure}

The magnetic susceptibility and specific heat are measured to
characterize our crystals. It has been known that the main feature
of $\chi(T)$ for pure LiCu$_2$O$_2$ is a broad maximum at $\sim$
36 K, which is a characteristic of quasi-one-dimensional magnet,
and the onsets of long-range magnetic orders at $T_{N1}$ and
$T_{N2}$ induce a sharp anomaly at the temperature derivative of
the magnetic susceptibility, $d\chi(T)/dT$, for magnetic field
along the $c$ axis and the $ab$ plane, respectively.\cite{Masuda}
These results are well reproduced in our crystals, as shown in
Figs. 2(a) and 2(b). Furthermore, upon doping Zn, the peak
positions of $d\chi(T)/dT$ gradually move to lower temperatures,
indicating the suppression of long-range magnetic order with
increasing nonmagnetic impurities.\cite{Hsu1, Hsu2} Figure 2(c)
shows the low-temperature specific heat of pure and Zn-doped
LiCu$_{2-x}$Zn$_x$O$_2$ single crystals. For pure sample, the
$C(T)$ curve shows two small but clear peaks at $T_{N1}$ = 24.7 K
and $T_{N2}$ = 23.2 K, respectively, which are known to be due to
the magnetic phase transitions from paramagnetic state to a
sinusoidal spin ordering and then to a helicoidal spin
ordering.\cite{Seki, Huang, Rusydi, Capogna, Moskvin} Upon doping
Zn, these two peaks shift to lower temperatures and become weaker
and their positions have good correspondence with those in
$d\chi(T)/dT$. When $x$ arrives 0.177, the two peaks in $C(T)$ are
not distinguishable from each other and evolute to a hump-like
anomaly at $\sim$ 17 K. This evolution of the specific heat with
Zn doping is compatible with some earlier reports,\cite{Hsu1,
Hsu2} in which it was discussed that Zn doping results in either
the phase transition of short range or significant inhomogeneity.

\begin{figure}
\includegraphics[clip,width=7.5cm]{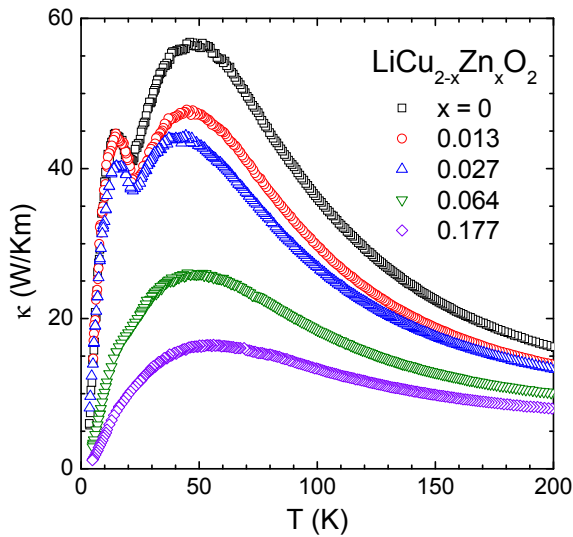}
\caption{(Color online) Thermal conductivity of pure and Zn-doped
LiCu$_2$O$_2$ single crystals.}
\end{figure}

Figure 3 shows the low-temperature thermal conductivity of pure
and Zn-doped LiCu$_2$O$_2$ single crystals. The thermal
conductivity of pure crystal is rather large but its temperature
dependence is apparently different from that of usual
insulators.\cite{Berman} With lowering temperature, the $\kappa$
increases quickly and achieves a high value of 57 W/Km at 48 K. A
remarkable feature of $\kappa(T)$ is the appearance of two peaks
at 48 K and 14 K. It is notable that the minimum between two peaks
is located at $\sim$ 22 K, having a good correspondence to the
positions of specific-heat peaks. This suggests a possible origin
of the double peaks of $\kappa(T)$, that is, their appearance is
due to the formation of a minimum, caused by strong phonon
scattering by the spin fluctuations at the critical regions of
magnetic transitions.\cite{Dixon1} This phenomenon was also found
in some multiferroic materials exhibiting strong spin-phonon
coupling, such as HoMnO$_3$.\cite{Wang} Although this explanation
does not bring obvious contradict with the impurity (Zn) doping
effect, however, it meets with difficulty in understanding the
magnetic-field dependence of thermal conductivity, as shown below.

It is found that the two $\kappa(T)$ peaks show rather different
dependences on the concentration of Zn. While the high-$T$ peak
weakens gradually with increasing $x$, the low-$T$ one shows a
weak $x$ dependence for low doping levels $x \le$ 0.027 but it is
so strongly damped for $x \ge 0.064$ that it is almost smeared out
completely. Note that the evolution of low-$T$ peak seems to have
some direct relationship to the Zn-doping effect on the
specific-heat data. As can be seen in Fig. 2, the two
specific-heat peaks are also smeared out for $x \ge 0.064$, which
manifests that the long-range spin orderings are destroyed by such
high impurity dopings. This comparison naturally suggests that the
low-$T$ peak of $\kappa(T)$, which locates at temperature just
below the phase transition of long-range spin ordering, is likely
due to the heat transport by the magnons in the
antiferromagnetically ordered state.

As far as the high-$T$ peak of $\kappa(T)$ is concerned, one
possible origin is the phonon peak as in usual
insulators.\cite{Berman} The magnitude of phonon peak is known to
be strongly dependent on the impurities and point defects in
crystals. Apparently, the strong impurity dependence of high-$T$
peak of LiCu$_2$O$_2$ is compatible with such standard behavior.
However, one may note that the position of this ``phonon peak" is
located at a bit too high temperature, compared to many other
materials which presenting phonon peak below 20 K.\cite{Berman}
Therefore, we need to consider another possible origin of the
high-$T$ peak due to the magnetic excitations transporting heat,
as evidenced in many low-dimensional spin systems, such as
SrCuO$_2$, Sr$_2$CuO$_3$, CaCu$_2$O$_3$,
Sr$_{14}$Cu$_{24}$O$_{41}$, La$_2$CuO$_4$.\cite{Sologubenko2,
Kawamata, Hess2, Hess3, Sun_LCO, Sun_PLCCO, Berggold} In these
materials, the magnetic term of thermal conductivity can be much
larger than the phononic term and is also sensitive to the
impurities. In particular, the magnetic excitations heat transport
in quasi-one dimensional spin-1/2 materials was predicted to be
the ballistic type. One direct experimental evidence for the
ballistic transport was the non-magnetic impurity doping effect,
in which the mean free path of magnetic excitations was found to
be very close to the average distance between spin
defects.\cite{Kawamata} In this regard, the magnetic heat
transport in these low-dimensional spin systems is usually
obtained from the strong anisotropy of thermal
conductivity;\cite{Hess1, Sologubenko1} that is, one can get the
purely magnetic thermal conductivity by subtracting the phonon
term, which is estimated from the thermal conductivity
perpendicular to the one-dimensional spin chain or the
two-dimensional spin network. It will be interesting to get the
magnetic heat transport in LiCu$_2$O$_2$ by using the same method
and compare the Zn-doping effect with that in other spin-1/2 chain
compounds. However, because of the in-plane twin structures of
LiCu$_2$O$_2$ crystals, one cannot separate the heat transport
along the $a$ and the $b$ axes. In addition, the heat transport
along the $c$ axis is also found to be impossible to measure
reliably because of the easy cleavage of the LiCu$_2$O$_2$
crystals along the $ab$ plane. It is therefore not feasible for us
now to get the magnetic heat transport along the spin chains in
these LiCu$_2$O$_2$ crystals. An effective way to detwin the
LiCu$_2$O$_2$ crystals is called for the investigation on the
possible role of the magnetic excitations transporting heat.

\begin{figure}
\includegraphics[clip,width=7.0cm]{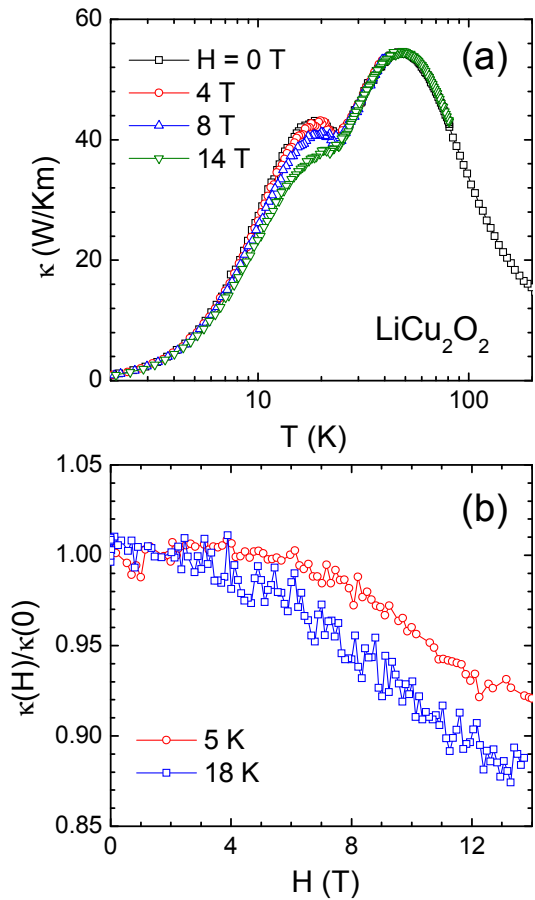}
\caption{(Color online) (a) Temperature dependences of thermal
conductivity of LiCu$_2$O$_2$ single crystal in 0--14 T. The
direction of magnetic field is along that of the heat current. (b)
Low-temperature $\kappa(H)$ isotherms at 5 K and 18 K.}
\end{figure}

The effect of magnetic field on the thermal conductivity is
studied for in-plane fields up to 14 T. First of all, as shown in
Fig. 4(a), the high-$T$ peak is completely independent of the
magnetic field, which would not be unreasonable if the high-$T$
peak is a pure phononic behavior. For some well-studied
low-dimensional magnetic materials, it is also found in recent
experiments that the magnetic heat transport is insensitive to the
external magnetic field, which is however due to the large
exchange coupling in these materials, typically being of the order
of magnitude of 100 meV.\cite{Hess3, Brenig2} The exchange
coupling in LiCu$_2$O$_2$ is known to be more than one order of
magnitude smaller\cite{Masuda} and is therefore not much larger
than the energy caused by the magnetic field in order of 10 T. In
this sense, the insensitivity of the high-$T$ thermal conductivity
to the magnetic field may not support the conjecture that the
magnetic excitations are responsible for transporting heat above
the magnetic phase transitions and the formation of high-$T$ peak.
Second, the ``dip" between two $\kappa(T)$ peaks does not show any
change in applied magnetic field, which immediately rule out the
possibility that the ``dip" is caused by strong spin-phonon
scattering. In contrast, it has already been found in many AF
materials that the strong magnetic field can suppress the spin
fluctuations and recover the thermal conductivity if the
spin-phonon coupling is considerably strong.\cite{Wang}

On the other hand, the low-$T$ peak is gradually suppressed by the
magnetic field, similar to those in many antiferromagnetically
ordered materials.\cite{Dixon2} The detailed field dependences of
$\kappa$ are shown in Fig. 4(b), in which two $\kappa(H)$
isotherms at 5 and 18 K are included. It is clear that the thermal
conductivity is monotonically decreased with increasing field,
without showing any signature of saturation or drastic transition
up to 14 T. This kind of field dependence is expectable for the
magnon heat transport since the magnons tend to be less populated
with increasing magnetic field. In addition, the field dependence
at 18 K, which is near the low-$T$ peak of $\kappa(T)$, is
stronger than that at 5 K. It is also understandable because the
magnetic heat conductivity is apparently much larger at 18 K. This
result shows a clear evidence that below the long-range-order
transition temperature the magnons act as heat carriers in
LiCu$_2$O$_2$. Note that both the $\kappa(T)$ and $\kappa(H)$
behaviors indicate that the magnon heat transport seems to be
weakened with lowering temperature. This is mainly due to the
decrease of magnon population and is reasonable for LiCu$_2$O$_2$
since there is a 1.4 meV gap in the magnetic excitation spectrum,
found by the electron spin resonance experiments.\cite{Mihaly} It
can be seen that the heat transport properties of LiCu$_2$O$_2$
are rather conventional,\cite{Dixon2} without showing any peculiar
behavior of the low-dimensional quantum magnets. The reason is
likely related to the rather strong spin frustration in this
material.

It has been known that the in-plane magnetic field can rotate the
spin directions and produce some spin-flop-like transitions at 2
T.\cite{Park} However, the $\kappa(H)$ curves do not show any
anomaly across these transitions, in contrast to some observations
in other compounds.\cite{Spin_flop, Wang, Zhao} One reason for the
drastic change of $\kappa$ at the spin-flop transition is that the
magnons are significantly populated because of the closure of the
anisotropy gap.\cite{Spin_flop} Therefore, it is not clear whether
the magnon spectrum is gapless at the spin re-orientation
transition in this spirally ordered antiferromagnet.

\section{Summary}

The Zn-doping and magnetic-field dependences of thermal
conductivity of LiCu$_2$O$_2$ single crystals are carefully
studied. The $\kappa(T)$ data show a double-peak phenomenon. The
higher-$T$ peak at 48 K is due to the phonon heat transport, while
the lower-$T$ peak at 14 K is a result of magnon heat transport
showing up in the magnetic long-range-ordered state. The present
results indicate that the magnetic heat transport of LiCu$_2$O$_2$
behaves similarly as that in the three-dimensional
antiferromagnets. The absence of the characteristic transport
properties of that in the low-dimensional quantum spin systems may
be related to the complexity of spin structure and spin
frustration.

\begin{acknowledgements}

This work was supported by the Chinese Academy of Sciences, the
National Natural Science Foundation of China, the National Basic
Research Program of China (Grants No. 2009CB929502 and
2011CBA00111), and the RFDP (Grant No. 20070358076).

\end{acknowledgements}

\end{document}